\documentclass[a4paper]{jpconf}
\usepackage{graphicx}
\begin{document}
\title{Superfluid gap formation in a fermionic optical lattice with spin imbalanced populations}

\author{Akihisa Koga$^1$, Philipp Werner$^2$}

\address{$^1$Department of Physics, Tokyo Institute of Technology, Tokyo 152-8551, Japan}
\address{$^2$Theoretische Physik, ETH Zurich, 8093 Z\"urich, Switzerland}

\ead{koga@phys.titech.ac.jp}

\begin{abstract}
We investigate the attractive Hubbard model in infinite spatial dimensions 
at quarter filling. 
By combining dynamical mean-field theory with 
continuous-time quantum Monte Carlo simulations in the Nambu formalism, 
we directly deal with the superfluid phase in the population imbalanced system. 
We discuss the low energy properties in the polarized superfluid state and the pseudogap 
behavior in the vicinity of the critical temperature.
\end{abstract}

\section{Introduction}
Recently, ultracold fermions have attracted much interest.
One of the most interesting topics in the field is 
the superfluid state~\cite{Regal},
where Cooper pairs composed of distinct components condense 
at low temperatures.
Owing to the high controllability of the interaction strength and 
the particle number, 
interesting phenomena have been observed such as 
the BCS-BEC crossover~\cite{BCSBEC1,BCSBEC2,BCSBEC3} and 
the superfluid state in imbalanced systems~\cite{Imbalance1,Imbalance2}.
These observations stimulate further experimental and theoretical investigations
on fermionic systems.

For optical lattice systems, 
which are formed by loading ultracold fermions into a periodic potential, 
various types of superfluid states have been proposed~\cite{1D,Chen,Koga}.
One of the possible candidates is the polarized superfluid state, 
which is naively expected to be realized 
in higher dimensional systems with imbalanced populations~\cite{ImbalanceDao}.
In our previous paper~\cite{KogaQMC}, 
we have investigated the attractive Hubbard model 
in infinite spatial dimensions to clarify how the polarized superfluid state 
is realized at low temperatures.
However, dynamical properties were not yet studied in detail 
in the intermediate correlation and temperature region. 
In particular, the question whether or not 
a pseudogap structure appears in the density of states
above the critical temperature, has not been answered.
To clarify this point, we study the formation of the superfluid gap
in the system with and without spin-imbalanced populations.

The paper is organized as follows.
In \S 2, we introduce the model Hamiltonian and 
briefly summarize the DMFT framework. 
In \S 3, we study the attractive Hubbard model at quarter filling.
By combining dynamical mean-field theory
 (DMFT)~\cite{Metzner,Muller,Georges,Pruschke}
with the continuous time quantum Monte Carlo (CTQMC) method~\cite{Rubtsov},
we discuss how the polarized superfluid state is realized and 
the superfluid gap appears in the density of states at low temperatures.
A brief summary is given in \S4.

\section{Model and Method}

We consider a correlated fermion system with attractive interactions,
which may be described by the Hubbard Hamiltonian,
\begin{equation}
\hat{\cal{H}}=\sum_{(i,j),\sigma}
\left[-t-\left(\mu+h\sigma\right)\delta_{ij}\right]c^{\dagger}_{i\sigma}c_{j\sigma}
-U\sum_{i}n_{i\uparrow}n_{i\downarrow},
\label{eq1}
\end{equation}
where $c_{i\sigma}$ ($c^{\dagger}_{i\sigma}$) 
is an annihilation (creation) operator of a fermion on the $i$th site
with spin $\sigma$, and 
$n_{i\sigma}= c^{\dagger}_{i\sigma}c_{i\sigma}$. 
$U$ is the onsite attractive interaction, $t$ is the transfer integral
between sites, $\mu$ is the chemical potential, and $h$ is the magnetic field,
which controls the spin imbalanced populations in the system.

To study the attractive Hubbard model in infinite spatial dimensions~\cite{ImbalanceDao,KogaQMC,Keller,Garg,Toschi,Bauer}, 
we make use of DMFT~\cite{Metzner,Muller,Georges,Pruschke}.
In DMFT, the original lattice model is mapped to an effective impurity model, 
which accurately takes into account local particle correlations.
The lattice Green's function is obtained via a self-consistency 
condition imposed on the impurity problem.
This treatment is formally exact in infinite spatial dimensions, and 
DMFT has successfully been applied to 
strongly correlated fermion systems.
When the superfluid state is directly treated in the framework of DMFT, 
the Green's function should be described 
by a $2\times 2$ matrix.
The self-consistency condition for the model~\cite{GeorgesZ} is given as,
\begin{equation}
\hat{G}_{0}^{-1}(i\omega_n) = \left(i\omega_n +h \right)\hat{\sigma}_0
+\mu\hat{\sigma}_z-\left(\frac{D}{4}\right)^2\hat{\sigma}_z\hat{G}(i\omega_n)\hat{\sigma}_z\label{eq:self},
\end{equation}
where $\hat{G}_0(i\omega_n) [\hat{G}(i\omega_n)]$ is 
the non-interacting (interacting) Green's function for the impurity model, 
$\omega_n [=(2n+1)\pi T]$ is the Matsubara frequency, $T$ is the temperature,
$\hat{\sigma}_0$ is the identity matrix, 
and $\hat{\sigma}_z$ is the $z$-component of the Pauli matrix. 
Here, we have used the semi-circular density of states, $\rho_0(x) = 2/\pi D \sqrt{1-(x/D)^2}$, where $D$ is the half bandwidth. 
In the following, we set $D$ as a unit of the energy.

When low energy properties in strongly correlated systems are discussed
in the framework of DMFT, an impurity solver is necessary 
to obtain the Green's function for the effective impurity model.
Here we use the recently developed CTQMC technique, 
which has successfully been applied to 
the Hubbard model~\cite{KogaQMC,CTQMC,Multi}, 
the periodic Anderson model~\cite{Luitz},
the Kondo lattice model~\cite{Otsuki}
and the Holstein-Hubbard model~\cite{Phonon}.
In this paper, by using a CTQMC method 
in the continuous-time auxiliary field formulation \cite{Gull}
and extended to the Nambu formalism as an impurity solver, 
we discuss how the polarized superfluid state is realized in the system.
The details of the method have been explained in Ref. \cite{KogaQMC}.

\section{Results}
\begin{figure}[h]
\begin{center}
\includegraphics[width=15cm]{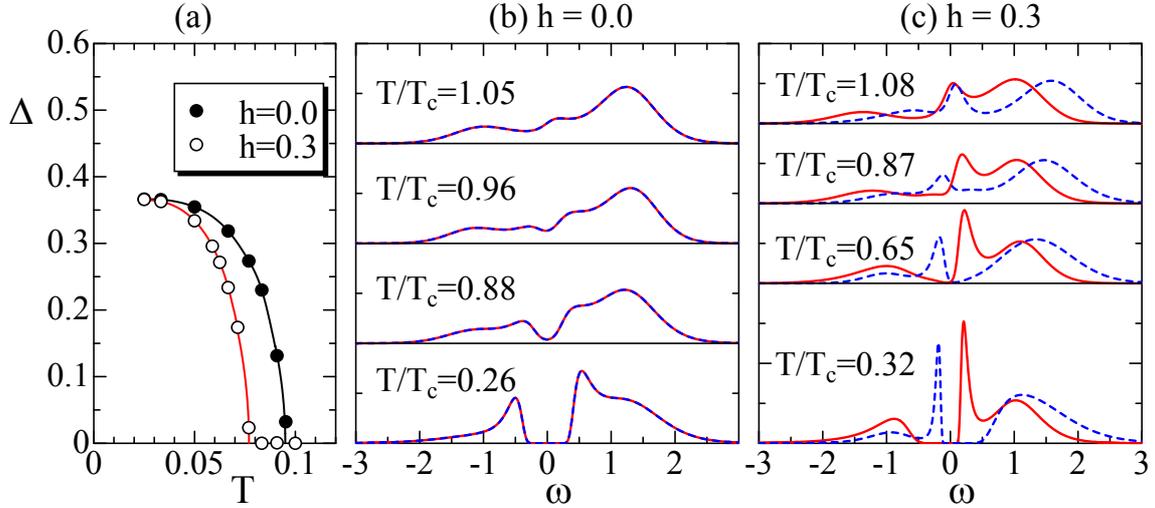}\hspace{2pc}%
\end{center}
\caption{
(a) Pair potential as a function of the temperature when $h=0.0$ and $h=0.3$.
Solid and dashed lines represent the spectral functions for fermions with up (down) spin when $h=0.0$ (b) and $h=0.3$ (c).}
\label{fig1}
\end{figure}

We study the superfluid state in the Hubbard model at quarter filling
to discuss how the gap structure appears in the density of states.
A quarter filled system is chosen to study the attractive Hubbard model at a generic filling.
The pair potential $\Delta (=\langle c_{i\uparrow} c_{i\downarrow}\rangle)$
and the density of states $\rho_\sigma(\omega)$ 
in the case $U=2$ are shown in Fig. \ref{fig1}.
When $h=0.0$, as temperature is decreased, 
the pair potential is induced at $T_c\sim 0.095$, 
as shown in Fig. \ref{fig1} (a).
This implies that a phase transition occurs 
from the normal metallic state to the superfluid state.
A superfluid gap characteristic of this state
clearly appears in the density of states at low temperatures, 
as shown in Fig. \ref{fig1} (b).
When the magnetic field is switched on, 
pairing correlations are suppressed and the normal metallic state becomes stable.
In fact, the superfluid order parameter is decreased 
and the critical temperature is shifted to $T_c\sim 0.077$
when $h=0.3$, 
as shown in Fig. \ref{fig1} (a).
When a large magnetic field is applied, the polarized superfluid state is no longer stable even at very low temperatures. This instability has been discussed in Ref. \cite{KogaQMC}.
The density of states for each spin 
in the $h=0.3$ case 
is shown in Fig. \ref{fig1} (c).
It is found that the density of states around the Fermi level decreases, 
with decrease of temperature from $T_c$,
similar to the case with $h=0.0$.
These results suggest that the superfluid gap appears only 
at $T<T_c$.
Therefore, we conclude that no pseudogap behavior is found  
at $T>T_c$,
at least, in the quarter filled Hubbard model with the interation $U=2$.
However, there still exists an open question whether or not a pseudogap behavior
appears in other parameter regions 
such as the weak or strong coupling region and other fillings.
Further systematic calculations are necessary to clarify this point,
and will be discussed elsewhere.

\section{Summary}
We have investigated the attractive Hubbard model 
in infinite spatial dimensions, 
by combining dynamical mean-field theory with continuous-time quantum 
Monte Carlo simulations based on the Nambu formalism.
We have calculated the superfluid order parameter
and the density of states to discuss 
how the polarized superfluid state is realized at low temperatures.
It was found that no pseudogap behavior appears above 
the critical temperature. 

\section*{Acknowledgment}
Parts of the computations were done on TSUBAME
Grid Cluster at the Global Scientific Information and Computing
Center of the Tokyo Institute of Technology. 
This work was partly supported by the Grant-in-Aid for Scientific Research 
20740194 (A.K.) and 
the Global COE Program ``Nanoscience and Quantum Physics" from 
the Ministry of Education, Culture, Sports, Science and Technology (MEXT) 
of Japan. PW acknowledges support from SNF Grant PP002-118866.


\section*{References}

\begin{thebibliography}{9}

\bibitem{Regal}
Regal C A, Greiner M, and Jin D S
2004 {\it Phys. Rev. Lett.} {\bf 92} 040403

\bibitem{BCSBEC1}
Jochim S, 
Bartenstein M, Altmeyer A, Hendl G, Riedl S, Chin C, Denschlag J H, 
and Grimm R 
2003 {\it Science} {\bf 302} 2101

\bibitem{BCSBEC2}
Zwierlein M W, Stan C A, Schunck C H, Raupach S M F, Gupta S, Hadzibabic Z, 
and Ketterle W 
2003 {\it Phys. Rev. Lett.} {\bf 91} 250401

\bibitem{BCSBEC3}
Bourdel T,
Khaykovich L, Cubizolles J, Zhang J, Chevy F, Teichmann M, Tarruell L, 
Kokkelmans S J J M F, and Salomon C
2004 {\it Phys. Rev. Lett.} {\bf 93} 050401

\bibitem{Imbalance1} 
Zwierlein M W, Schirotzek A, Shunck C H, and Ketterle W
2006 {\it Science} {\bf 311} 492

\bibitem{Imbalance2}
Partridge G B, Li W, Kamar R I, Liao Y, and Hulet R G 
2006 {\it Science} {\bf 311} 503



\bibitem{1D}
Pour F K, Rigol M, Wessel S, and Muramatsu A, 
2007 {\it Phys. Rev. B} {\bf 75} 161104;
Xianlong G, Rizzi M, Polini M, Fazio R, Tosi M P, 
Campo Jr. V L, and Capelle K 2007 {\it Phys. Rev. Lett.} {\bf 98} 030404;
Tezuka M and Ueda M 2008 {\it Phys. Rev. Lett.} {\bf 100} 110403; 
Machida M, Yamada S, Okumura M, Ohashi Y, and Matsumoto H 
2008 {\it Phys. Rev. A} {\bf 77} 053614

\bibitem{Chen}
Koponen T K, Paananen T, Martikainen J P, and T\"orm\"a P 
2007 {\it Phys. Rev. Lett.} {\bf 99} 120403;
Chen Y, Wang Z D, Zhang F C, and Ting C S 
2009 {\it Phys. Rev. B} {\bf 79} 054512; 
Tamaki H, Miyake K, and Ohashi Y 
2009 {\it J. Phys. Soc. Jpn.} {\bf 78} 073001

\bibitem{Koga}
Koga A, Higashiyama T, Inaba K, Suga S, and Kawakami N  
2008 {\it J. Phys. Soc. Jpn.} {\bf 77} 073602; 
2009 {\it Phys. Rev. A} {\bf 79} 013607;
Fujihara Y, Koga A, and Kawakami N, 
2009 {\it Phys. Rev. A} {\bf 79} 013610


\bibitem{ImbalanceDao}
Dao T L, Ferrero M, Georges A, Capone M, and Parcollet O
2008 {\it Phys. Rev. Lett.} {\bf 101} 236405

\bibitem{KogaQMC}
Koga A and Werner P 2010 {\it J. Phys. Soc. Jpn.} {\bf 79} 064401

\bibitem{Metzner}
Metzner W and Vollhardt D 
1989 {\it Phys. Rev. Lett.} {\bf 62} 324

\bibitem{Muller}
M\"uller-Hartmann E 
1989 {\it Z. Phys. B} {\bf 74} 507

\bibitem{Georges}
Georges A, Kotliar G, Krauth W and Rozenberg M J
1996 {\it Rev. Mod. Phys.} {\bf 68} 13

\bibitem{Pruschke}
Pruschke T, Jarrell M, and Freericks J K
1995 {\it Adv. Phys.} {\bf 42} 187

\bibitem{Rubtsov}
Rubtsov A N, Savkin V V and Lichtenstein A I 
2005 {\it Phys. Rev. B} {\bf 72} 035122

\bibitem{Keller}
Keller M, Metzner W, and Schollw\"ock U
2001 {\it Phys. Rev. Lett.} {\bf 86} 4612


\bibitem{Garg}
Garg A, Krishnamurthy H R, and Randeria M 
2005 {\it Phys. Rev. B} {\bf 72} 024517

\bibitem{Toschi}
Toschi A, Capone M, and Castellani C
2005 {\it Phys. Rev. B} {\bf 72} 235118

\bibitem{Bauer}
Bauer J, Hewson A C, and Dupuis N 
2009 {\it Phys. Rev. B} {\bf 79} 214518;
Bauer J and Hewson A C
2009 {\it Europhys. Lett.} {\bf 85} 27001


\bibitem{GeorgesZ}
Georges A, Kotliar G, and Krauth W
1993 {\it Z. Phys. B} {\bf 92} 313


\bibitem{CTQMC}
Werner P, Comanac A, de'Medici L, Troyer M, and Millis A J
2006 {\it Phys. Rev. Lett.} {\bf 97} 076405;
Werner P and Millis A J
2007 {\it Phys. Rev. B} {\bf 75} 085108


\bibitem{Multi}
Werner P and Millis A J
2006 {\it Phys. Rev. B} {\bf 74} 155107; 
2007 {\it Phys. Rev. Lett.} {\bf 99} 126405;
Werner P, Gull E, and Millis A J
2009 {\it Phys. Rev. B} {\bf 79} 115119

\bibitem{Luitz}
Luitz D J and Assaad F F
2010 {\it Phys. Rev. B} {\bf 81} 024509


\bibitem{Otsuki}
Otsuki J, Kusunose H, Werner P, and Kuramoto Y
2007 {\it J. Phys. Soc. Jpn.} {\bf 76} 114707; 
Otsuki J, Kusunose H, and Kuramoto Y
2009 {\it Phys. Rev. Lett.} {\bf 102} 017202

\bibitem{Phonon}
Assaad F F and Lang T C 
2007 {\it Phys. Rev. B} {\bf 76} 035116;
Werner P and Millis A J
2007 {\it Phys. Rev. Lett.} {\bf 99} 146404

\bibitem{Gull}
Gull E, Werner P, Parcollet O and Troyer M
2008 {\it Europhys. Lett.} {\bf 82} 57003





\end{thebibliography}
\end{document}